\documentclass[12pt,a4paper,final]{iopart}

\usepackage{iopams}
\usepackage{array}
\usepackage{cite}
\usepackage{graphicx}
\usepackage[breaklinks=true,colorlinks=true,linkcolor=blue,urlcolor=blue,citecolor=blue]{hyperref}

\begin{document}

\title[Thermal modulation for suppression of parametric instability in advanced gravitational wave detectors
]{Thermal modulation for suppression of parametric instability in advanced gravitational wave detectors}

 \author{Y.B. Ma$^{1,2,3}$, J. Liu$^3$, Y.Q. Ma$^3$, C. Zhao$^3$, L. Ju$^3$, D.G. Blair$^3$, Z.H. Zhu$^1$}
\address{$^1$ Department of Astronomy, BeiJing Normal University, BeiJing, 100875, China }
\address{$^2$ School of Physic, ShanXi DaTong University, Datong, 037009, China}
\address{$^3$ School of Physic, The University of Western Australia,Crawley, WA 6009, Australia}
\ead{yuboma.phy@gmail.com}

\begin{abstract}
Three-mode parametric instability is a threat to attaining design power levels in Advanced gravitational wave detectors. The first observation of three-mode parametric instability in a long optical cavity revealed that instabilities could be suppressed by time variation of the mirror radius of curvature. In this paper, we present three dimensional finite element analysis of this thermo-acousto-optics system to determine whether thermal modulation could provide sufficient instability's suppression without degrading time averaged optical performance. It is shown that deformations due to the time averaged heating profile on the mirror surface can be compensated by rear surface heating of the test mass. Results show that a $CO_2$ laser heating beam with a modulation amplitude of $1$ Watt at 0.01 Hz is sufficient to stabilize acoustic mode with parametric gain up to 3. The parametric gain suppression factor is linearly proportional to the peak modulation power.

\end{abstract}

%\maketitle

\section{Introduction}

Laser interferometer gravitational wave detectors are the most sensitive displacement sensors that human-beings have ever built. They use laser light to sense the test mass motion induced by passing gravitational waves.
%The working mechanism of these detectors are based on opto-mechanical interactions \cite{ref1}, that is, the test masses motion driven by the radiation pressure force and the gravitational waves.
For reaching high sensitivity, one has to increase
%the optomechanical interaction strength, which is typically depended on
the optical power that interacts with test masses to reduce the quantum shot noise. Advanced detectors utilise Fabry-Perot cavities in the interferometer arms, combined with power recycling, to aim to achieve a circulating optical power inside the cavity approaching 1 MW.

However, such a high optical power on the test mass has undesired effects, which include thermal aberrations \cite{Lawrence2004,Hello2006} due to unavoidable optical absorption of the test masses, and instabilities induced by radiation pressure \cite{Braginsky2001,Sidles2006}. One particular effect that we are interested in this paper is the three-mode parametric instability.
%inside the Fabry-Perot cavities.
The coupling of the high power optical field with the internal acoustic vibration of the test mass mirrors can lead to uncontrolled optomechanical oscillations which leads to instability. Parametric instability limits the circulating power, thereby limiting attainable detector sensitivity.

Three-mode parametric instability involves one acoustic mode and two optical modes. It was first predicted in a seminal work by Braginsky \textit{et al}. in 2001 \cite{Braginsky2001, Braginsky2002} and then investigated in detail using 3D finite element simulation for Advanced LIGO gravitational wave detectors \cite{Zhao2005,Gras2010}.  Three-mode parametric instability in free space Fabry-Perot optomechanical cavities was first observed in a desktop optical experiment in 2013 \cite{Chen2015} and later observed in an $80$ m suspended cavity, in which LIGO \textbf{type} conditions were mimicked, at the Gingin High Optical Power Facility in Western Australia \cite{Zhao2015}. In 2014 parametric instability was observed in an Advanced LIGO detector at LIGO Livingston Observatory (LLO) \cite{Evans2015}. An acoustic mode with $Q = 12\times10^{6}$ and resonant frequency of 15.5 kHz was observed to undergo exponential growth with threshold circulating power of 25kW, finally causing the interferometer lose lock.  The observation was in a good agreement with predictions \cite{Gras2010}.

Various methods were proposed for controlling parametric instability \cite{Zhao2005,Vyatchanin2012,Evans2010,Ju2009,Degallaix2007, Miller2011,Gras2009,Gras2015,Evans2008}. Because parametric gain is a function of the difference between the optical cavity mode gap between the fundamental mode and higher order transverse mode, and the acoustic mode frequency, tuning of the optical mode gap through thermal tuning of the mirror radius of curvature proved to be an effective way of suppressing parametric instability.  However, due to the high acoustic mode density of the test masses, simple thermal tuning would have limited ranges. At the Gingin 80 m suspended optical cavity, an interesting effect was observed: the amplitude growth of an acoustic mode due to parametric instability was modulated \cite{Zhao2015}, leading to an effective suppression of parametric gain. It was shown that this effect was due to radius of curvature modulation. The modulation occurred because the cavity optical transverse mode frequency was modulated due to the low frequency residual fluctuations of the beam spot position. Beam spot position in turn modulated the transverse mode frequencies due to figure errors of the mirror which cause every spot position to "see" a different radius of curvature. In a system where figure errors of the mirrors are very small, the radius of curvature variation with time could be provided by thermal modulation.
% It was also pointed out that this mechanism of dynamic modulation suppression of parametric gain could be used as a new method to control instability.
% Because parametric gain is a function of the difference between the optical cavity mode gap and the acoustic mode frequency, modulation of the optical mode gap through thermal modulation of the mirror radius of curvature could provide appreciable instability suppression.
%from the transverse mode frequency modulation. The modulation in this experiment was due to figure errors of the mirror and residual fluctuations of the beam spot position. It was also pointed out that this mechanism of dynamic modulation suppression opens the new method to control instability. It was proposed that modulation of the optical mode gap through thermal modulation of the mirror radius of curvature  could provide appreciable instability suppression. Because parametric gain is a function of the difference between the optical cavity mode gap and the acoustic mode frequency, modulation causes time dependent detuning. An ideal thermal modulation system was simulated
This idea was preliminary explored using an ideal thermal modulation model in ref. \cite{Zhao2015}.
%with heating/cooling capability (where it is practically difficult to achieve) to show the idea.
%The thermal modulation causes differential expansion, which acts to create a local change in radius of curvature.

%In this paper, using the Advanced LIGO parameters, we are going to investigate the feasibility of a practical thermal modulation system for modulating the radius of curvature and thereby the cavity mode gap for suppressing the parametric instability
%: to use $CO_2$  laser to heat the mirror surface of Advanced LIGO detector.
In this paper, we investigate the feasibility of a practical thermal modulation scheme, using the Advanced LIGO parameters, for suppressing the parametric instability. Using 3D finite modeling analysis, we show that thermal modulation of the mirror, combined with back side static heating, could provide sufficient parametric instability suppression without degrading time averaged optical performance.  The paper is organized as follows: in section \ref{sec2_PIestimation}, we give a theoretical background of this work and orders of magnitude estimation; Section \ref{sec3_sim} presents the simulation results, in which section \ref{sec3_1} establishes the simulation mode and discusses the static effect that needs to be considered in our scheme. Section \ref{sec3_2} presents numerical results of the dynamical modulation and the suppression of the instability.
%The paper is organized as follows: in section 2, we give a theoretical background of this work and orders of magnitude estimation; Section 3 presents our simulation result, in which the first subsection establishes our simulation mode and discuss the static effect that need to be considered in our scheme; while the second subsection presents numerical results of the dynamical modulation and the suppression of the instability. At the end we give the conclusion of this paper.

\section{The principle and orders of magnitude estimation}\label{sec2_PIestimation}

In an optical cavity, with suspended test mass mirrors, the  thermal fluctuations drive large numbers of acoustic modes. Each acoustic mode scatters the intra-cavity fundamental mode field (frequency $\omega_{0}$) into high order transverse optical modes.  There will be beating between the fundamental optical mode and a high order optical mode (frequency $\omega_{1}$), with beat frequency $\Delta\omega = \omega_{0}-\omega_{1}$. If $\Delta\omega$ coincides with the frequency of the acoustic mode of the mirror $\omega_{m}$, and if the mode shape of the high order optical mode and acoustic mode shape have a certain level of overlap, parametric opto-acoustic interaction will occur via radiation pressure associated with the beating. If $\omega_{0}-\omega_{1}> 0$, the acoustic mode will be excited, leading to parametric instability. The parametric gain can be expressed as \cite{Ju2006}:

\begin{equation}\label{eq1_R}
 R=\frac{P\Lambda\omega_{1}}{M\omega_{m}L^{2}\gamma_{m}\gamma_{0}\gamma_{1}}
 \frac{1}{1+(\Delta_{m}/\gamma_{1})^{2}},
\end{equation}
where $M$ is the mass of the mirror, $L$ is the arm cavity length, $\Lambda$ is the overlapping factor of the optical mode and acoustic mode, $\gamma_{m}$, $\gamma_{0}$, $\gamma_{1}$ are the relaxation rate (half-linewidth) of the acoustic mode and the two optical modes respectively. $\Delta_m =(\omega_0-\omega_1) - \omega_m$ is the frequency detuning factor which determines the detuning of the cavity mode spacing from the  test mass acoustic mode frequency. When $R>1$, the radiation pressure induced ring-up rate $\gamma_{opt}$ is lager than the intrinsic relaxation rate of the acoustic mode $\gamma_{m}$, then the system is unstable.

%It is clear that the fundamental mode power $P$ and the overlapping factor $\Lambda$ which measures the matching between the spatial distribution of the optical field and the mirror surface vibration are positively correlated with the rate of the mechanical amplitude increasing, $\gamma_{opt}$, since these quantities determine the pondermotive coupling strength. However, if the frequency of the scattered light is detuned from the resonant peak of the high order optical mode, then apparently the parametric gain will be reduced. Therefore the $\gamma_{opt}$ should be negatively correlated with the ratio of the detuning $\Delta_{m}=(\omega_{0}-\omega_{1})-\omega_{m}$ to the cavity high order mode relaxation rate $\gamma_{1}$.
%These scaling relation is demonstrated in the formula of the parametric gain $R=\gamma_{opt}/\gamma_{m}$, which is
A critical factor in determine parametric instability is the optical cavity mode gap $\Delta\omega=\omega_0-\omega_1$, which depends directly on the radii of curvature of the cavity mirrors $R_1$ and $R_2$.  The cavity mode spacing is given by
% As we can see from Eq.(\ref{eq1_R}), the parametric gain depends on the frequency detuning $\Delta_{m}$ between the acoustic frequency and the optical cavity mode spacing   $\Delta\omega=\omega_0-\omega_1$, which is given by:

\begin{equation}
  \Delta\omega=\frac{c}{L}(m+n)\cos^{-1}\left[\pm\sqrt{(1-\frac{L}{R_{1}})(1-\frac{L}{R_{2}})}\right],
\end{equation}
where $m$, $n$ are the transverse mode indies, and $R_{1}$ and $R_{2}$ are the radii of curvature of the input and end mirrors of the cavity.

If one of the test mass mirrors (eg. end mirror) is heated up due to absorption of  optical power (from the main cavity mode, or from a separate $CO_2$ laser beam applied to it), thermal distortion of the mirror will modify its radius of curvature $R_{2}\rightarrow R_{2}+\delta R_{2}$.  The mode spacing can be written (expanded to the first order) as:

\begin{equation}\label{eq3}
\Delta\omega^{(1)}\approx\Delta\omega^{(0)}+\beta\delta R_{2}.
\end{equation}
Here,
\begin{equation}\label{eq4}
\beta=(m+n)L(L-R_{1})/2R_{2}\sqrt{(L(L-R_{1})(L-R_{2})(R_{1}+R_{2}-L))},
\end{equation}
while $\Delta\omega^{(0)}$ is the original mode spacing before heating.

As a simple example, assume a laser beam with radius $w$ and power $P_{0}$ is applied to a spherical mirror of radius of curvature $R_{roc}$. Due to absorption and finite thermal conductivity of the mirror, a temperature difference between the central part and the edge of the mirror will develop within its substrate \cite{Winkler1991}:
\begin{equation}
\delta T\approx P_{abs}/2\pi\kappa w,
\end{equation}
where $P_{obs}$ is the absorbed power at the centre and $\kappa$ is the thermal conductivity of the mirror material.  If the material has a thermal expansion coefficient of $\alpha$, the thermal distortion expressed as a change of sagitta $\delta d$, and the consequent change of mirror RoC are given by \cite{Winkler1991}:
\begin{equation}
\delta d=\alpha w\delta T/2\approx\alpha
P_{abs}/(4\pi\kappa),
\end{equation}
and
\begin{equation}
  \delta R_{roc}=(w/d)^{2}\delta d/2,
\end{equation}
where $d$ is the initial sagitta with the cord length equal to the heating beam size. For fused silica, $\alpha\sim5.5\times10^{-7}$ m/K, $\kappa\sim1.38$ W/m.K. An absorbed power $\sim1$ W will lead to $\delta d\sim 30$ nm.  For mirrors designed to have figure errors of $\sim 0.5$ nm , this is an enormous effect. The deformation corresponds to a mode-spacing change of the order of $\sim 2000$ Hz. This would be enough to tune $\Delta_{m}$ by several optical mode linewidths and $\sim 10^{3}$ acoustic linewidths; and for LIGO type test masses, to tune across $\sim 10^{2}$ acoustic modes from parametric instability condition.

Thermal tuning based on this method but using radiant heating was  used in the advanced LIGO detectors to control parametric instability for an input power of $\sim 1/4$ of the designed power \cite{Evans2015}.  However, this simple thermal tuning method may not be sufficient if there are multiple potential unstable acoustic modes present, because tuning the cavity away from one unstable acoustic mode normally leads it to be tuned into resonance with another acoustic mode.

% \textbf{
% For parametric gain $\sim 3$, the ring up time constant is typically $\sim$ 1 minute \cite{Ju2006-1}. For example, the amplitude of a 20 kHz acoustic mode with $Q = 10^7$ and $R = 3$, the amplitude  will grow 5 orders higher in 15 min. MAYBE THIS IS NOT NECESSARY}
We now consider the application of a periodic $CO_2$ laser thermal modulation $P = P_0(1+\sin\Omega t)$. The period $2\pi/\Omega$ is chosen to be much larger than the cavity build up time $\tau\sim4L/(cT_r)$ ($T_r$ being the transmissivity of the input mirror of the cavity), but much smaller than the typical time for exponential growth of parametric instability. The modulation will cause the cavity to sweep through unstable resonant conditions while never remaining within a resonant condition long enough for  acoustic unstable modes to build up. The time scale of the modulation $2\pi/\Omega$, typically can have the value from $1-100$ s which is smaller than the thermal conduction time of fused silica, $t_{th}\sim l^{2}c\rho/\kappa\sim10^{4}s $,  where $l$ is the typical length scale (test mass radius of 0.17m), $c$ is the specific heat and $\rho$ is the density.
%These qualitative analysis implies the time-dependence of the temperature of the mirror surface as the following.

The modulation power has both a constant (DC) and sinusoidal (AC) component.  The corresponding DC temperature of the surface will increase to a certain value when the thermal dissipation balances the power absorption. The thermodynamic process under the periodic heat modulation is non-equilibrium due to the slow thermal response of the material. Because the test masses are suspended by thin wires in the vacuum, the dominant heat dissipation is by radiation. The radiated power can be estimated using the Stefan-Boltzmann law $\sigma l^{2}(T^{4}-T^4_{0})$, where $\sigma$ is the Stefan-Boltzmann constant, $T$ and $T_{0}$ are the temperatures of the test mass surface and the environment respectively. For a power absorption of $\sim 20$W, one can estimate the final temperature to be $\delta T\sim P_{abs}/4\sigma l^{2}T^{3}_{0}\sim8$ K, which implies that the $CO_2$  heating does not have a significant effect on the thermal noise level. Use of the Stefan-Boltzmann law is justified because fused silica is effectively a black body at thermal wavelengths.  With a modulated heating power, the AC temperature at the heating spot position will vary with the heating power variation. However, the thermal conduction will result in a phase lag of the temperature variation to the heating power variation. The amplitude of the temperature variation decreases as the heating power modulation frequency increases, as does the variation of the radius of curvature and the mode spacing.

\section{Simulation results}\label{sec3_sim}

We used the finite element modeling software package, COMSOL, to perform a thermal elastic analysis of the test mass mirrors. We used a cylindrical fused silica test mass model. Some detailed features of the real test mass for the advanced gravitational wave detectors, such as the flat surfaces on the barrel for suspension, are neglected for simplicity, since they are far from the laser heating center. Fused silica has very high absorption at the $CO_2$ laser wavelength of  $10\mu m$, the optical power is mostly absorbed at the heating spot surface with a penetration depth of micrometer scale. We treated the heated surface as the heat source which transports energy into the substrate and radiates energy into the environment. Due to the small absorption depth of the fused silica, the highest temperature area will be close to the substrate surface. We used a geometric sequence mesh to ensure enough resolution near the surface. The $CO_2$ laser intensity distribution over the light spot is modeled as a two-dimensional Gaussian distribution with a sinusoidal time-dependence: $P(t) = P_0(1 + \sin \Omega t)$. For studying the cavity mode spacing change and optical mode wavefront deformation of the Fabry-Parot cavity under the in uence of the heating, we used optical Fast Fourier Transformation code, OSCAR \cite{Degallaix2010}. The parameters that used in the simulation is given in Table \ref{tab1}.

\begin{table}
\caption{\label{tab1}The main parameters used for the simulation}
\footnotesize
\centering
\begin{tabular}{@{}ll}
\br
Parameters&Value\\
\mr
Test mass material                                         &\verb Fused    silica\\
Test mass size                                             &\verb 34cm(diam.) $\times$ 20cm\\
Arm cavity length                                          &\verb 3994.5 m\\
Arm cavity finesse                                         &\verb 450      \\
Radius of curvature, ITM / ETM                             &\verb 1934 m / 2245 m\\
Modulation Laser Power on HR surface                       &\verb 1W, $CO_{2}$ \\
Modulation Laser Beam radius ($1/e^2$), ETM                &\verb 6 cm\\
Compensation Laser Power on AR surface                     &\verb Annular, 17W, $CO_{2}$ \\
\br
\end{tabular}\\
\end{table}
\normalsize

\subsection{Static effects}\label{sec3_1}

\begin{figure}[htb!]
\centering
\includegraphics[width=0.6\linewidth]{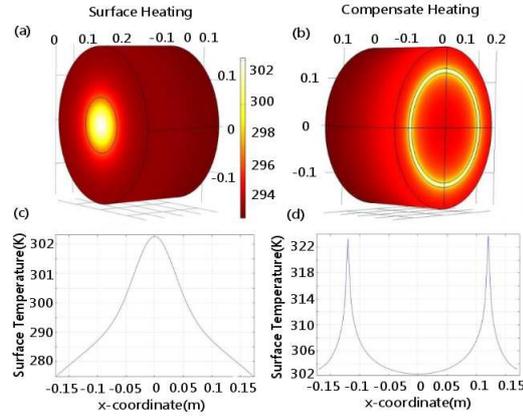}
\caption{\label{figureone}Heating scheme: (a) and (b) show the temperature distribution on the front and back surfaces of the test mass; the bright yellow spot on the front surface and the ring on the back surface show the heating beam pattern.
%describes the how the DC part of the heating beam and the compensating heating beam change the surface temperature distribution of whole mirror.
(c) and (d) are the temperature distribution along the diameter of the front and back surfaces.
%of the high reflecting surface and the AR surface, respectively.
The power of the heating beam and the compensation beam are  1 W and 17 W.
%As we can see from these figure, the thermal noise increase due to the heating is not a significant issue for a room-temperature detector.
}
\end{figure}

We will first examine the time averaged, effectively static effects of the thermal modulation scheme to determine how it affects the operation of the optical cavities. With the modulated heating power mentioned above, there would be a non-zero average heating applied on the surface and thus a static thermal deformation of the mirror.
%effect generated purely by the  $CO_2$  laser Gaussian beam will cause a static (DC) distortion of the mirror, which basically modify the boundary conditions of the electromagnetic fields inside the cavity, thereby create the coupling between different cavity modes.
This deformation will change the cavity mode shape and cause
%Especially, there will be
a mismatch between the input light and the hot cavity mode, resulting undesirable extra losses. To overcome this problem, we applied an additional heating beam to the back surface of
the test mass to compensate the deformation created by the the front surface heating. Figure \ref{figureone} shows the temperature profile of this compensation scheme. The Gaussian
beam heating at the center of the front surface creates a bulge at the center, while the ring type heating on the back surface induces test mass bending towards the front surface.

The mismatch due to the mirror surface distortion can be quantified using the mode matching factor:

\begin{equation}
\eta=\frac{\int_{S}\psi_{0}(x)\psi^*_{1}(x)d^{2}x}{\sqrt{\int_{s}|\psi_{0}(x)|^{2}d^{2}x}\sqrt{\int_{s}|\psi_{1}(x)|^{2}d^{2}x}}
\end{equation}
where $\psi_{1}(x)$ and $\psi_{01}(x)$ are the spatial field distribution of the cavity mode with and without heating distortion, respectively.

\begin{figure}[htb!]
\centering
\includegraphics[width=0.6\linewidth]{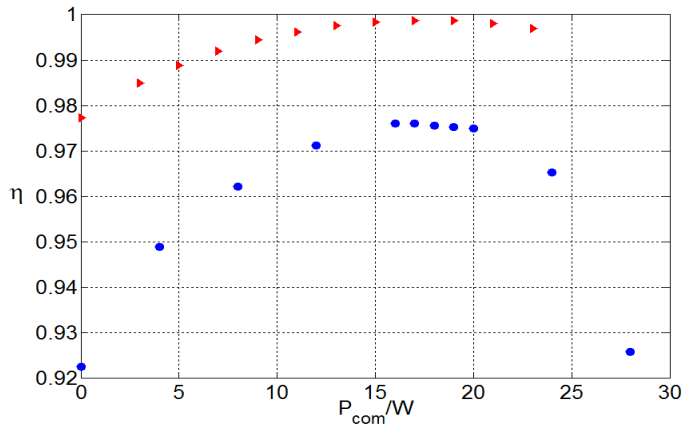}
\caption{\label{figuretwo} The mode matching factor $\eta$ changes with different back surface compensation power. The top and bottom curves correspond to 6cm and 3cm front surface heating spot radius with 1 W heating power, respectively.}
\end{figure}

Figure \ref{figuretwo} shows the mode matching factor changes as a function of the back surface compensation heating power, with a constant 1W heating power on the front surface. It is clear that there exists an optimum compensation heating power on the back surface, i.e. 17W in this particular case.
% Figure \ref{figurethree} shows the front surfaces deformation with a 6cm, 1W front heating and different back surface heating powers.

%The change in the value of $\eta$ tells us how the mode match changes with respect to the change of the distortion of the mirror due to the heating of both the Gaussian $CO_2$  beam and the compensation beam. The Fig. \ref{figuretwo} shows the mode matching change as a function of the compensation power, while the heating power on the front surface is constant 1W. The Fig. \ref{figurethree} shows and the typical front surfaces deformation with front and back surface heating. It is clear that there exists a optimum compensation heating power applied on the back surface, i.e. 17W in this particular case.
%($TEM_{00}$ mode ) while the $\psi_{1}(x)$ describes the spatial field distribution of the distorted cavity mode after heating. The change in the value of $\eta$ tells us how the mode match changes with respect to the change of the distortion of the mirror due to the heating of both the Gaussian $CO_2$  beam and the compensation beam. The Fig. \ref{figuretwo} shows the mode matching change as a function of the compensation power, while the heating power on the front surface is constant 1W. The Fig. \ref{figurethree} shows and the typical front surfaces deformation with front and back surface heating. It is clear that there exists a optimum compensation heating power applied on the back surface, i.e. 17W in this particular case.

% \begin{figure}[htb!]
% \centering
% \includegraphics[width=0.5\linewidth]{2.png}
% \caption{\label{figurethree} Typical front surface deformation profile with both the front and back surfaces heating. Here the front heating  spot radius is 6cm and the different compensation power on the back surface are: 5 W, 9 W, 13 W, 17 W, and 23 W. The dashed vertical lines indicates the spot size of the front heating laser beam}
% \end{figure}

%From these two figures, it is clear that the value $\eta$ takes a maximum value at some intermediate compensation power. This fact can be physically understood by noticing that for the $\eta$ approaches 1, the strain distribution should have the following feature that (1) the distance between two peaks $d_{P}$ should be larger than the spot size of the carrier field for preventing the scattering loss and diffraction loss; (2) the radius of the curvature of the mirror surface region $R_{dis}$ covered by the carrier field must be close to the radius of the curvature of the wavefront. Our simulation shows that $d_{P}$P decreases with the compensation power increase while $R_{dis}$ increase.

From Fig. \ref{figuretwo} it can be seen that when the heating spot radius is 6cm, the back-side heating scheme can indeed compensate the static effect quite well with a compensation power of 17 W.  The mode matching factor $\eta$ reached $\sim 0.998$, or power mode matching efficiency of $\sim99.6\%$.  This corresponds to a $0.4\%$ power lost to high order modes, which is $\sim 8$ times smaller than power recycling mirror power transmission of LIGO design and should not decrease the power recycling gain much.
%which corresponds to a loss around 4000 ppm, smaller than the transmissivity of the input test mass.
While with the front heating spot radius of 3 cm, the maximum $\eta$ can only reach 0.976, or power mode matching efficiency of $93.2\%$.
%this corresponds to a much larger loss which is 5 times larger than the input mirror transmission.
This suggests that if the front modulation heating beam size is equal to the main cavity beam size, an optimised back surface compensation works well with high mode matching efficiency.

%%%%%%%%%%%%%%%%%%%%%%%%%%
\subsection{Dynamical modulation and the suppression of parametric instability}\label{sec3_2}

%Now consider the effect of the time-dependent modulating heating $P_0(1+\sin\Omega t)$.
Having evaluated the steady state effects we now go on to consider its effectiveness in suppressing instability. In section 3.1, we saw that after an initial temperature rise due to the front and back surface heating, the average temperature of the test mass reached a steady state. However the front surface deformation changes sinusoidally.

%Having discussed about the effect of static deformation on the mode shape, now we add the time-dependent sinusoidal part to the heating beam on the high reflective surface. Under the action of the sinusoidal heating power and the compensation beam, the thermal-elastic deformation of the mirror surface behaves in the way of sinusoidal in time.
%in which we choose the power of the compensation beam to be the optimal regarding to the mode shape distortion.

% \begin{figure}[htb!]
% \centering
% \includegraphics[width=3in,height=2in]{5.png}
% \caption{\label{figurefour} }
% \end{figure}

%Note that this figure shows the position of the center point of the mirror surface, which typically represents the time-dependent behavior of the deformation of the whole surface. It is clear that the time-  dependent deformation has a static (DC) component and also a time-dependent (AC) component, where the amplitude of the AC component
%around the DC distortion decreases when the modulation frequency increases. This comes from the fact that the lower modulation frequency (or longer period) allows the heating/cooling process closer to the adiabatic approximation.
%the material to follow more adiabatically on the change of the heating power.

\begin{figure}[htb!]
\centering
\includegraphics[width=0.6\linewidth]{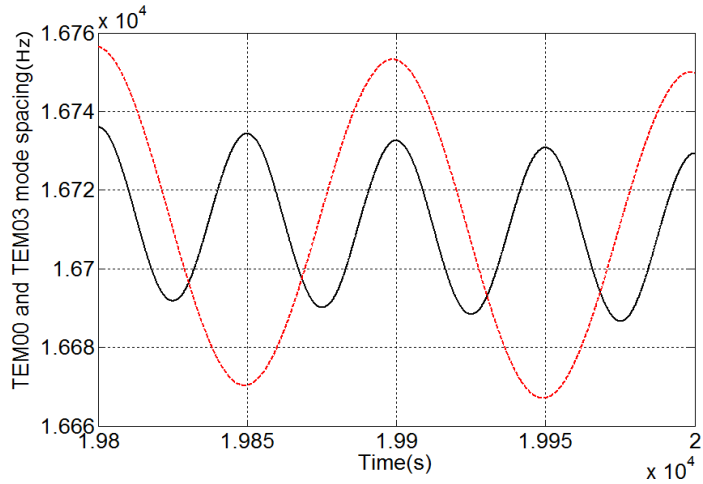}
\caption{\label{figurefour} Mode spacing between $TEM_{00}$ mode and $TEM_{03}$ mode with a front surface modulation $CO_2$ laser heating of 1 W and back surface compensation 17 W heating, as shown in Fig.\ref{figureone}. The black (solid) and red (dashed) curves correspond to modulation frequencies of $f=0.01$ Hz and 0.02 Hz, respectively.}
\end{figure}

With the deformation data of the test mass front surface, we simulate the optical cavity modes and their spacing using OSCAR.
%the mode spacing $\Delta\omega$ introduced in Eq.(2) using OSCAR,
Figure \ref{figurefour} shows the modulated cavity mode spacing between $TEM_{00}$ mode and $TEM_{03}$ mode, with $CO_2$ laser front and back heating at two modulation frequencies.  The mode gap is modulated by 10's of Hz, and as expected, the amplitude of the periodic change of the optical cavity mode spacing decreases as the modulation frequency increases.

%Note that in this figure,
%In Fig. \ref{figurefour}, we have only chosen a limited time window $1.98\times 10^4 $ s - $-2\times 10^4 $ s to show the amplitude of periodic change of optical cavity mode spacing, which decreases as the modulation frequency increases.
%The static mode spacing is actually increasing with time, but our simulation shows that the amplitude of the periodic change of parametric instability is almost a constant.
It was shown in reference \cite{Zhao2015} that the equivalent parametric gain $R_a$ could be reduced by modulating the optical cavity mode spacing:
%Substitute the mode spacing results into the parametric gain formula [14]:
\begin{equation}
\centering
R_{a}=\frac{R_{max}}{\sqrt{1+4 a^2}}
\end{equation}
where $R_{max}$ is the maximum parametric gain when the mode spacing is exactly equal to the acoustic mode frequency. Factor $a$ here is the ratio between the modulation amplitude of the mode spacing and the linewidth of the higher order optical mode of the optical cavity.

\begin{figure}[htb!]
\centering
\includegraphics[width=0.65\linewidth]{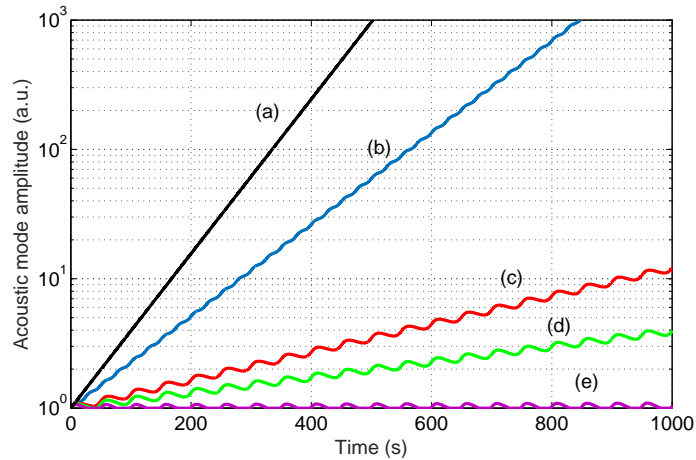}
\caption{\label{figurefive} Parametric instability suppression with different modulation heating frequency.  Curve $(a)$ shows the amplitude growth of an unstable acoustic modes with parametric gain of R=2.7.  Curves (b) and (c) are the 15.5 kHz acoustic mode amplitude as a function of time, with heat modulation frequencies of 20 mHz and 10 mHz respectively. Curves (d) and (e) are the 20.5kHz and 25kHz acoustic mode amplitudes as functions of time respectively, with heat modulation frequency of 10 mHz, while the initial parametric gains are the same, 2.7.  The vertical axis is the acoustic mode amplitude expressed in the units of its initial value.  The heating power is the same as in Fig. \ref{figurefour}.}
%The heating modulation actually slows down the amplitude increasing rate and  parametric gain without thermal modulation is 2.
\end{figure}

Figure \ref{figurefive} shows a comparison of instability suppression with different thermal modulation frequencies for three different acoustic modes that all have a parametric gain of 2.7 in the absence of dynamic modulation.  We modeled a 15.5 kHz unstable acoustic mode which matches to the 3rd order optical cavity mode, a 20.5 kHz acoustic mode which matches to a fourth order optical cavity mode, and a 20.5 kHz acoustic mode which matches to a fourth order optical cavity mode.   Curve (a) shows the 15.5 kHz unstable acoustic mode amplitude growth as a function of time in the absence of thermal modulation. Curves (b) and (c) show the same acoustic mode amplitude growth with thermal modulation frequencies of 20 mHz and 10 mHz respectively. Curves (d) shows the 20.5 kHz acoustic mode amplitude growth with thermal modulation frequency of 10 mHz. Curve (e) shows the 25 kHz acoustic mode amplitude growth with thermal modulation frequency of 10 mHz.

%Here a 1 Watt modulation heating power, as well as an assumption that all the optical modes have the same half linewidth of 45Hz were used.

%increasing of the mean square root of the acoustic mode amplitude is described in .

It can be seen that in the case of the 15.5 kHz acoustic mode (curves a, b and c), a thermal modulation amplitude of 1 Watt at frequency of 10 mHz significantly reduced the effective parametric gain, greatly slowing down the rate of acoustic amplitude growth.  This would allow more response time to apply other suppression methods such as electrostatic feedback damping before the interferometer loses lock. It can also be seen that thermal modulation can effectively suppress multiple acoustic modes simultaneously. Curves (c), (d) and (e) represent thermal modulation suppression effect on 3 acoustic modes (15.5kHz, 20.5kHz and 25kHz) with the same initial parametric gain of 2.7.  As shown in Eqs.\ref{eq3} and \ref{eq4}, with the same radius of curvature tuning, higher order optical modes have larger mode spacing tuning.  Therefore the higher frequency acoustic modes experience larger modulation amplitudes with the same thermal modulation. In the case of the 25kHz acoustic mode, with 0.1 Hz thermal modulation, the acoustic mode amplitude fluctuates around the original amplitude, resulting in an effective  parametric gain of $\sim 1$, completely suppressed the instability.

%%%%%%%%%%%%%%%%%%%%%%%
\section{Conclusions}\label{sect4-Conclusion}

We have undertaken a detailed simulation of an Advanced LIGO type optical cavity to test the effectiveness of thermal modulation of the test masses for the suppressing of parametric instability. We have shown that the method, as previously predicted, is capable of suppressing parametric instability for modes with parametric gain in the range 1-3. It has the advantage of suppressing multiple unstable modes simultaneously.
% Based on the numerical modeling of the advanced LIGO test mass mirror and the corresponding optical cavity using COMSOL and OSCAR, we studied parametric instability suppression using modulated $CO_2$ laser heating.
%We studied the effect of the laser heating to the wavefront of the main carrier laser beam in the arm cavity and find out the way to compensate this wave front distortion effect: by adding another 17W laser beam with ring power distribution on the anti-reecting surfaces.
Because single sided modulation results in a non-zero average heating power, leading to a steady state changes in RoC of a test mass and thereby a static cavity mode spacing changes, we proposed a ring type heater at the back surface of the test mass to compensate for the steady state effects. This would not be necessary if the test mass RoC is pre-designed to take into account the static heating induced RoC change.

Our results show multiple unstable modes with parametric gain of $2\sim 3$ can be suppressed by modulated heating with a power of 1 Watt.  For higher parametric gain, a higher modulation amplitude power is required to suppress the instability.
%We then numerically simulate the time-dependent dynamical heating effect, which shows that the observed parametric instability on advanced LIGO with parametric gain $R = 4$ could indeed be suppressed using our method.
Even if the parametric gain is too high for this scheme to completely suppress the instability, the thermal modulation can be used to increase the acoustic mode build-up time,
%slow down the increasing rate of the thermal amplitude of the acoustic mode can be slowed down
so that other control schemes can be implemented within a relatively long period of time. %and have small effect on the thermal noise of the detection sensitivity.

\section*{Acknowledgment}
The authors would like to thank Aidan Brooks for reviewing the draft. This work is supported by Australian Research Council and is part of the work of Australian Consortium for Gravitational wave Astronomy.This work was supported by the National Natural Science Foundation of China under Grants No. 11633001.

\end{document}